\documentclass[%
 reprint,
 amsmath,
 amssymb,
 aps,
prb,
]{revtex4-1}
\usepackage{graphicx}
\usepackage[T1]{fontenc}
\usepackage{epstopdf}
\usepackage{units}
\usepackage{dcolumn}
\usepackage{bm}

\begin{document}

\title{Anomalously large spin-current voltages on the surface of SmB$_6$}

\author{Johannes Geurs}
\author{Gihun Ryu}
\author{Chengtian Lin}
\author{Jurgen H. Smet}
 \email{j.smet@fkf.mpg.de}
\affiliation{Max Planck Institute for Solid State Research, Heisenbergstrasse 1, D-70569 Stuttgart}%

\date{\today}
\begin{abstract}
The spin-polarized surface states of topological insulators have attracted interest both from a fundamental and applied point of view. A recent proposal describes a method of probing these surface states with ferromagnetic contacts, which was subsequently applied to a variety of materials. In this study, we use this method on the potential topological insulator SmB$_6$ with a new design based on the Corbino geometry. Though the signal behaves as predicted for all orientations of current and magnetic field, its magnitude is much larger than expected. Possible parasitic effects such as stray field-induced Hall voltages are excluded, leaving the origin of the observations uncertain. This corroborates the need for careful analysis when interpreting results of similar experiments.
\end{abstract}

\maketitle

The prediction and realization of materials in which the topology of the electron wave function has physically observable effects has dominated condensed matter research in recent years. Most research has concentrated on topological insulators (TIs)\cite{Hasan2010d}, insulators of which the conductance and valence bands must cross at the interface with a topologically trivial material. The resulting conductive surface states exhibit a wealth of interesting properties, such as protection against backscattering and spin-momentum locking\cite{Ando2013}.

\begin{figure}
    \includegraphics[width=0.37\textwidth]{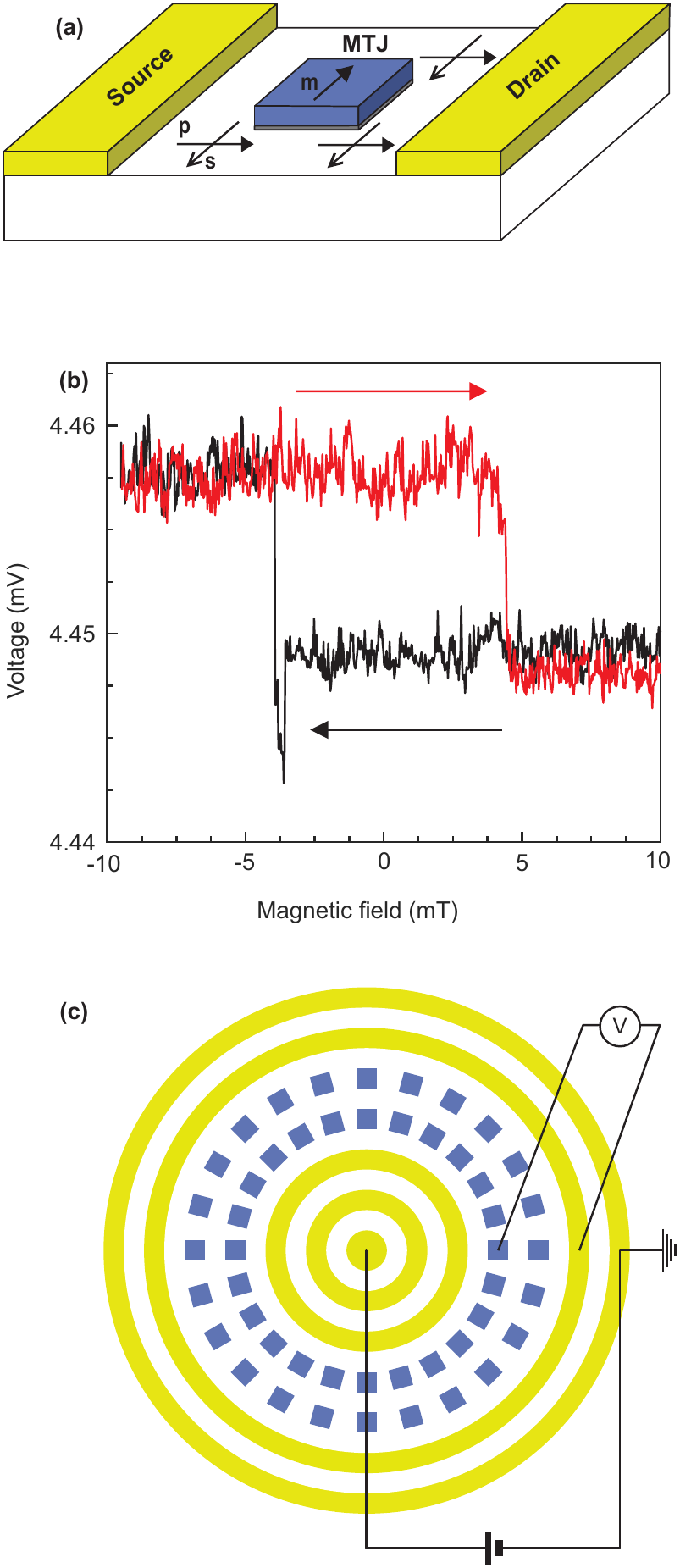}
    \caption{a. Potentiometric setup as described by Hong et al. \cite{Hong2012a} By applying a bias between the source and drain electrodes, charge current passes underneath a magnetic tunnel junction (MTJ). On the surface of a topological insulator, the spin \textbf{s} of a charge carrier must be perpendicular to its momentum \textbf{p}. If this current-induced spin polarization is aligned with the magnetization \textbf{m} of the MTJ, the potential at the MTJ is expected to increase. b. Voltage of the MTJ as a function of applied magnetic field. As the magnetic field is decreased from 10 mT to -10 mT (black trace), the magnetization of the MTJ flips at a coercive field of approximately -4 mT. This is accompanied by a sudden increase in voltage. On sweeping the magnetic field back (red trace), the voltage decreases suddenly at the positive coercive field, completing the loop. c. The Corbino geometry used in this experiment. Current passes radially outwards, and FM contacts are present at all orientations. The measured voltage of the MTJ is referenced to an annular gold contact close to ground. The total current that flows in the Corbino disk is $I_{DC}=10\mu A$.}
    \label{fig:1}
\end{figure}

Samarium hexaboride is possibly a member of this new class of materials. For many years\cite{Nickerson1971}, it has been known to increase in resistivity when cooled down. Uniquely, the increase saturates and reaches a plateau. This behaviour has been interpreted recently\cite{Dzero2010,Alexandrov2013} as that of a topological Kondo insulator: at low temperature, the bulk of the crystal becomes insulating and conduction can only take place over the surface. An ingenious transport experiment\cite{Wolgast2013} confirmed the current flowing over the surface at low temperatures.
The topological nature of this transition, however, is more difficult to prove experimentally.
In photoemission\cite{Xu2014}, the presence of surface bands has been proven, which are possibly spin-polarized. The detection of the small band gap of SmB$_6$ (10 meV) and in-gap states are however hindered by the energy resolution of this technique. De Haas-Van Alphen oscillations have been observed at high magnetic fields. Usually this is an excellent probe of the Fermi surface and Berry phase, but two studies have found two opposite interpretations: SmB$_6$ either has topological surface states\cite{Li2014h} or possesses a three dimensional Fermi surface\cite{Tan2015a,Hartstein2017}.
Electrical detection of spin polarization could resolve this issue. A recent work \cite{Hong2012a} proposed to use ferromagnetic (FM) tunnel contacts on the surface of TIs. The approach is schematically presented in Figure \ref{fig:1}a and described in the following.


A voltage bias is applied between two contacts on the surface of the TI, causing a charge current to flow in the direction {\textbf p} (Fig.~\ref{fig:1}a). Since the spin and momentum of a state must be locked and perpendicular to each other, that part of the current that flows over the surface is spin-polarized. This creates a difference in electrochemical potential between the different spin species. A third, ferromagnetic contact is separated by a tunnel junction from the TI surface. This magnetic tunnel junction (MTJ) has a spin-dependent tunnel resistance. With the constraint that no current flows out of the FM contact, its potential (the open-circuit voltage over the MTJ) must adapt to reflect the spin polarization underneath \cite{Hong2012a,Appelbaum2016}.
When the magnetization $\textbf{m}$ of the contact is flipped, its potential is predicted to change by
\begin{equation}
    \Delta V = \frac{h}{e^2}\cdot\eta I\cdot\frac{1}{k_FW}\cdot\cos\theta
    \label{eq:1}
\end{equation}
where $\eta$ is the degree of spin polarization of the surface current and $k_FW$ is an estimate of the number of channels underneath the contact, with the magnitude of the Fermi wave vector  $k_F=\sqrt{n/2\pi}$ and $W$ the width of the contact. In Eq.~(1) $h$ is Planck's constant, $e$ is the elementary charge, $I$ is the current flowing underneath the contact, and $\theta$ is the angle between the magnetization $\textbf{m}$ of the FM contact and the spin polarization $\textbf{s}$ of the surface current.


In this study, we investigate the behaviour of such MTJs on the surface of SmB$_6$ in an attempt to address the open question whether an imposed current can indeed generate a non-zero spin polarization of the surface states.
The temperature induced topological transition SmB$_6$ undergoes, may serve as a convenient tool to monitor the disappearance or emergence of these topological surface states. Hysteretic voltages over the MTJ are observed and do follow every symmetry predicted by Eq.~\ref{eq:1}, including the expected behaviour upon in-plane rotation of the magnetization of the FM contact. Hence, at first sight this confirms the creation of a current-induced spin polarization of topological origin. However, the magnitude $\Delta V$ of the signal by far exceeds every prediction, which casts doubt on the origin of the observed signal. A closer examination of the role of the ferromagnetic contact allows us to demonstrate that parasitic effects are unavoidable with this technique. However, the predicted magnitude of these parasitic effects is found to be even smaller than that of the spin-current induced voltage. Hence, the origin of the observed signals remains unaccounted for.

Our studies were performed on large single-crystal pieces (up to 8 mm) of SmB$_6$. The crystals were obtained using Xenon floating zone technique. The samarium hexaboride has a cubic crystal structure and cannot be exfoliated like typical TIs\cite{Kim2013e}, but cleaving is possible to obtain a fresh surface. If this crystal preparation proceeded in an oxygen free environment, the mobility reached values up to 140 ${\rm cm}^2/{\rm Vs}$ and the carrier density was typically around $10^{12}$ ${\rm cm}^{-2}$. Unfortunately, for the measurements described in this work, surface polishing was required to fabricate properly working tunnel junctions. This treatment has a detrimental impact on the quality of the surface and increases the carrier density to beyond $10^{14}$ ${\rm cm}^{-2}$. Even if so, no subsurface cracks were detected with electron microscopy. 
The crystals were aligned and cut in the (100) direction and polished to sub-nm roughness with progressively smaller diamond abrasive. In a second step, magnetic tunnel junctions were prepared by sputtering a 1.2 nm thick layer of TiO$_2$ followed by 85 nm of cobalt and 3 nm of Al as a capping layer. The size of the  tunnel junctions was $80 {\rm \mu m} \times  80 {\rm \mu m}$ in order to be able to make electrical connections by gluing bond wires onto the aluminium capping layer.
The tunnel barrier serves multiple purposes\cite{Dash2009c}.  For one, it protects the topological insulator from magnetic contamination, which would locally destroy the surface states.  It also counteracts the conductivity mismatch between the ferromagnetic contact and the ${\rm SmB}_{6}$ substrate, thereby enhancing the signal.
The sample was cooled down in a dilution refrigerator to the base temperature ($\sim$20 mK). Measurements were initially performed with DC current. They were subsequently repeated with AC biasing to confirm the results.


The fabrication of a conventional Hall bar geometry is problematic. The entire surface of the three dimensional substrate is conductive and can not be selectively depleted to create a Hall bar, so the direction and amplitude of the current are essentially unknown at any point on the surface. Hence, a Corbino-like geometry has been used instead, as illustrated schematically in Fig.~\ref{fig:1}c. A bias is applied between the central contact and the outermost ring, both made out of gold. Charge current flows radially outwards. Square magnetic tunnel junctions are positioned on top of this current-carrying channel. The potential of a FM contact is measured in a four-terminal configuration. It is referenced against another circular contact not carrying any external current and close to ground. A high-impedance preamplifier ($> 10G\Omega$) is used to prevent current flow through the magnetic contact. Fig.~\ref{fig:1}b illustrates a typical measurement. An external in-plane magnetic field controls the contact magnetization of the tunnel junction and is oriented perpendicular to the current flow underneath the junction. When this field is large enough, the contact magnetization will align with the field. If now a spin-polarized current flows underneath, the voltage will be modified. When the external field is gradually lowered and reversed (black curve), the contact magnetization will not change until the coercive field (around -4 mT, usual for thin cobalt films\cite{Munford2002}) is reached. At that point, the magnetization suddenly reverses. The spin and magnetization are now aligned and the potential at the contact of the tunnel junction is raised. When sweeping back the magnetic field (red curve), the original situation is recovered. The total change in voltage, $\Delta V$, is predicted by Eq.~\ref{eq:1}.

\begin{figure}
    \includegraphics[width=0.4\textwidth]{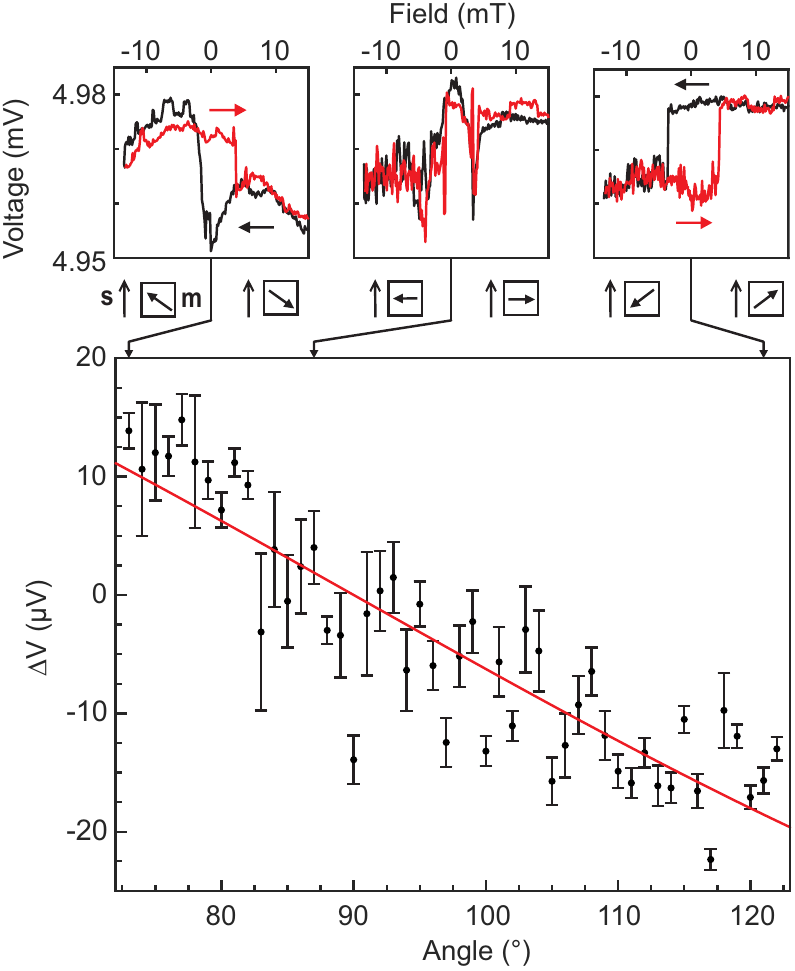}
    \caption{Influence of the magnetization of the ferromagnetic contact. Equation 1 predicts $\Delta V_S$ to depend on the angle between the contact magnetization and spin, $\theta$, which is shown by the red line. This is confirmed when rotating the contact magnetization: the magnitude of the hysteresis shrinks, goes to zero and then flips when past $\textbf{m}\cdot\textbf{s}=0$. Around 90 degrees, the switching is more erratic. Total current is $I_{AC}=14.1\mu$A$_{rms}$. The smaller panels show the switching behaviour at three different angles, with diagrams of the orientations of contact magnetization $\textbf{m}$ and spin polarization $\textbf{s}$.}
    \label{fig:2}
\end{figure}

The Corbino geometry uniquely allows for rotating the in-plane  magnetization of the ferromagnetic contact in the plane of the disk. Since these contacts are large and square, the magnitude of the coercive field will hardly change, only its direction. Fig.~\ref{fig:2} summarizes the results of such a rotation experiment. The external field is still swept to positive and negative values, but now the field axis is rotated and hence the magnetization will point away from \textbf{s}. As this axis approaches \textbf{p}, $\Delta V$ shrinks as predicted by Eq.~\ref{eq:1}. When the magnetization direction is parallel to the electrical current ({\bf m} $\|$ {\bf p}), i.e. perpendicular to the spin polarization ({\bf m} $\perp$ {\bf s}), $\Delta V$ reaches zero. The potential is not featureless in this case , but exhibits a random switching pattern. We attribute it to domain dynamics within the ferromagnetic contact. When the magnetization is rotated past $\textbf{s}$, a non-zero $\Delta V$ develops again but with opposite sign.

\begin{figure}
    \includegraphics[width=0.4\textwidth]{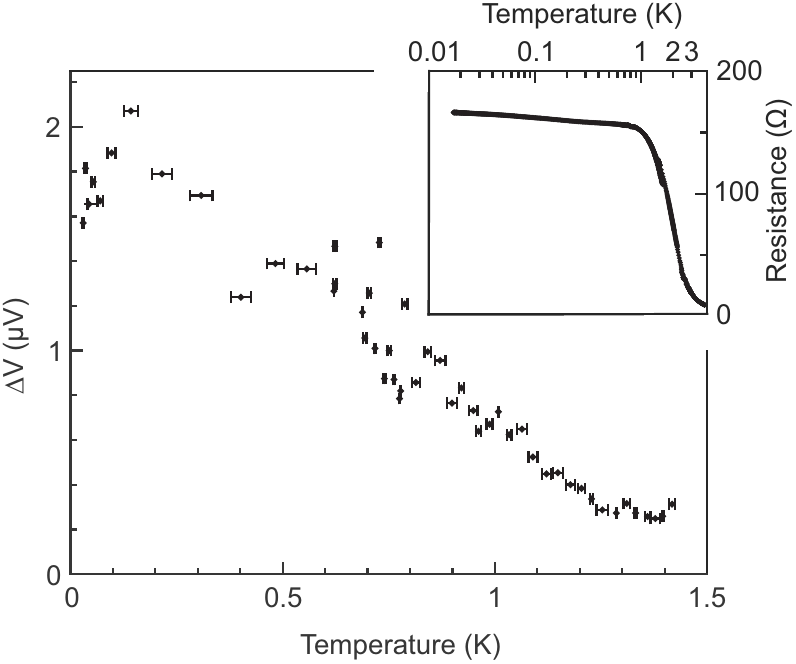}
    \caption{Temperature dependence. SmB$_6$ is suspected to go through a topological transition at high temperatures. $\Delta V$ decreases quickly and vanishes around 1.5K. Inset: the resistivity of SmB$_6$ features a plateau at low temperatures, where the entire current flows over the surface. The total current amounts to $I_{DC}=10.6\mu A$.}
    \label{fig:3}
\end{figure}

The saturation of the resistance of SmB$_6$ at low temperatures as observed in the inset to Fig.~\ref{fig:3} has been interpreted as the regime where conduction only takes place over the surface\cite{Neupane2013,Wolgast2013}.
If the measured signal were to persist beyond this temperature, topological states would need to be excluded as the origin of the observed effect. Fig. \ref{fig:3} shows the temperature dependence of $\Delta V$. As the temperature increases, $\Delta V$ decays quickly and has disappeared before the topological transition temperature is reached and current flows through the bulk. The fast weakening of the signal is reminiscent of the behaviour of counterdoped TIs\cite{Tang2014a}, tuned to have an insulating bulk. In these cases, the temperature dependence has been interpreted as either thermally activated bulk conduction or increased scattering at higher temperatures.

In the experimental configuration described here, Eq.~\ref{eq:1} predicts a voltage step of 150 nV, even when assuming the entire current flows at the surface and is completely spin polarized ($\eta =1$). For the more realistic case of partial polarization the expected signal scales down accordingly. The typical signal amplitude observed in experiment is however a few $\mu V$. This discrepancy of more than one order of magnitude between the predicted amplitude and experiment calls for a different interpretation of the experiments. In the following paragraphs, we will consider several possible parasitic effects.


The Rashba effect, where strong spin-orbit coupling splits the bands of a semiconductor in two spin-polarized copies, is a possible topologically trivial origin of spin-polarized current. Rashba splitting has been observed in bismuth selenide\cite{King2011}, but has been excluded as the origin of spin-current signals\cite{Li2014b} because the spin polarization created by Rashba splitting points in the opposite direction of the spin polarization created by spin-momentum locking in topological surface states. This argument can however not be applied to SmB$_6$ to exclude spin orbit coupling splitting as the origin of the observed signal, because in this material the large pockets at the X points\cite{Xu2014} have a spin texture that creates spin polarization in the same direction the Rashba splitting would. Even if so, the magnitude of the signal created by the Rashba effect would be far too small to explain the observations. Not only is the Rashba coupling of SmB$_6$ much smaller\cite{Hlawenka2015} than that of Bi$_2$Se$_3$, it has been shown on general grounds that the magnitude of the spin polarization created by the Rashba effect is always smaller than that created by spin-momentum locking in topological surface states\cite{Hong2012a}.

\begin{figure}
    \includegraphics[width=0.35\textwidth]{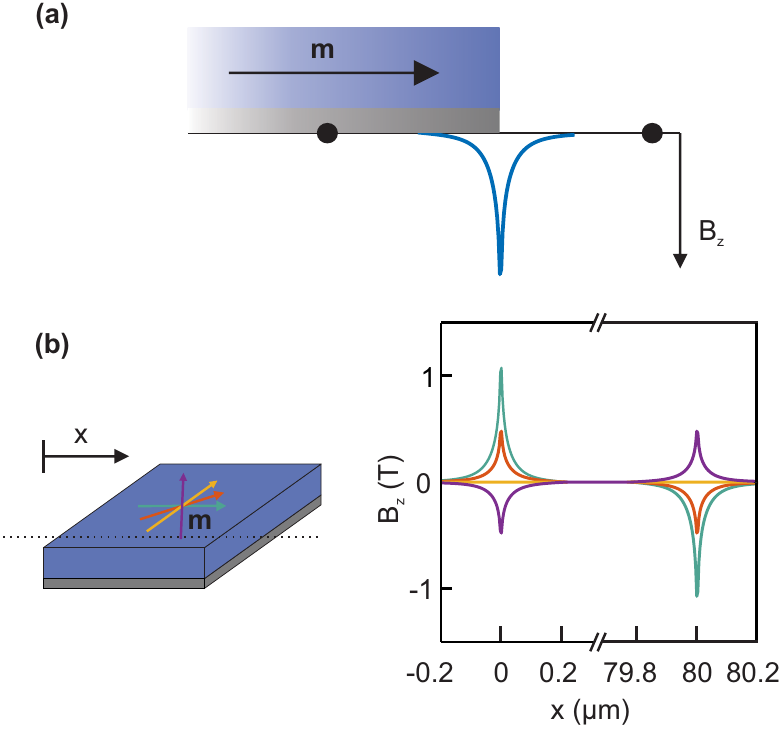}
    \caption{a. The fringe field (vertical component B$_z$ shown) surrounding half of a ferromagnetic contact (blue with magnetization \textbf{m}) pierces the surface states underneath (black line). If a current flows over the surface, underneath the FM contact, a Hall voltage will develop between a point below the FM contact and a point outside. b. Simulation of the magnetic field profile along the dashed line in the left schematic. When rotating the contact magnetization (coloured arrows), the stray field changes, locally reaching values up to 1 T. For four representative cases, the vertical component of the stray field underneath the MTJ is shown.}
    \label{fig:4}
\end{figure}

Recent studies\cite{DeVries2015c,Appelbaum2016} have suggested that this hysteretic signal can also arise from the stray magnetic field of the FM contact itself.
Stray fields of the ferromagnetic contacts pierce the surface near the contact boundaries as illustrated in Fig.~\ref{fig:4}a. The electrical current coming from the central contact is deflected by the perpendicular component of these stray fields. A Hall voltage will develop to counteract the deflection of the current and the local potential of the substrate underneath the ferromagnetic contact will be modified. The potential of the contact itself will follow this change.
If the magnetization of the contact is flipped, its stray fields and the developed Hall voltage will change sign. The potential of the contact will then change by an amount $\Delta V_H$.
We can estimate the $\Delta V_H$ of this parasitic effect by considering an infinite conductive sheet instead of our circular geometry. The carrier density of the sheet is assumed to be \textit{n} and the current density equals \textit{j}. Locally, a magnetic field \textit{B} pierces a ribbon along the current direction. The Hall voltage that develops across the ribbon is then equal to $\frac{1}{ne}\cdot \int jBdx$, where the integral runs across the ribbon.

The same argument holds for the geometry in this experiment: the local potential of the substrate underneath the FM contact will be modified by the same amount. The change in voltage between two opposite contact magnetization directions is then approximated as
\begin{equation}
    \Delta V_H\thicksim\frac{2}{ne}\cdot \int jB_zdx,
    \label{eq:2}
\end{equation}
integrating from a point underneath the contact to a point outside. This relation is not exact, since the ferromagnetic contact could quench some of the Hall voltage developing in the substrate underneath.
Further imperfections, such as deviations in the shape of the contact and roughness of the tunnel barrier, strain-induced effects and spreading of the current because of the Corbino geometry make equation \ref{eq:2} an estimate of the observed signal. It also assumes any changes in the potential underneath the contact are reflected in the potential of the contact itself. The treatment of this parasitic effect bears some resemblance to a related experiment\cite{Geim1997}.
Simulation of the FM contact in micromagnetic simulation software \cite{Donahue1999} shows that its demagnetization field can reach over 1 T at the edge.
Under rotation of the magnetization of the contact, the quantity $\int B_zdx$ will diminish as the magnetization direction becomes more parallel to the current, shown in Fig.~\ref{fig:4}b. Our simulation confirms that this decrease follows a $\cos(\theta)$ dependence. The angular dependence of $\Delta V_H$ is therefore identical to $\Delta V_S$ (Eq. 1), which makes it impossible to distinguish between the effects.

The temperature dependence plotted in Fig.~\ref{fig:3} can also be interpreted in this framework. When the temperature is increased, the current moves from the surface to the bulk. The magnetic fields that penetrate the surface and cause a local Hall effect no longer can have such a large effect; the vast majority of the bulk current is not deflected by a local magnetic field. The decrease in $\Delta V$ therefore reflects the decrease in the proportion of current that flows over the surface.
In the ideal case, $\Delta V_H$ as calculated above would be on the order of 5 $nV$, even smaller than the value for spin-current induced voltages.

One last parasitic effect is generated when current flows through the ferromagnetic contact \cite{Ando2017}. If the surface states are more resistive than the ferromagnetic material, a fraction of the current can tunnel into the contact and create an anomalous Hall voltage between the top and bottom of the FM contact. For any reasonable values of the anomalous Hall coefficient \cite{Miyasato2007}, the magnitude of this effect should be below 1 $nV$.

Our investigation of the behaviour of ferromagnetic tunnel junctions on the surface of SmB$_6$ raises questions about the interpretation of these signals. The observed signals follow all symmetries predicted for spin-current induced voltages, such as current and contact magnetization direction, and disappear when the current moves from the surface into the bulk. This is consistent with the prediction that SmB$_6$ is a topological Kondo insulator. However, the observed signals are more than an order of magnitude larger than this mechanism would predict.
Analysis of parasitic effects, such as stray field-induced Hall voltages, shows that these effects should obey the same symmetries as the above mechanism. Their magnitude, however, is anticipated to be even smaller than that of spin-current induced effect. This leaves the interpretation of the observations uncertain, and stresses the need for careful analysis when examining the results of similar experiments.

\section*{Acknowledgments}

The authors wish to thank Barbara Baum, Annette Zechmeister, Yvonne Stuhlhofer and Marion Hagel for help during sample fabrication, and Alexander Hoyer, Johannes F{\"o}rster and Georg Dieterle for insightful discussions.

\bibliography{library}

\end{document}